\documentclass[onecolumn,twoside,10pt]{article}
\usepackage{graphicx}
\usepackage{floatfig}

\begin{document}
\initfloatingfigs
\title{Structural anisotropy of silica hydrogels prepared \\
under magnetic field}
\author{Atsushi Mori$^a$\thanks{Correspoinding anthor.
Tel: +81-88-656-9417, E-mail address: mori@opt.tokushima-u.ac.jp}
\and
Takamasa Kaito$^b$\thanks{present address: KRI Inc., Tel:+81-75-322-6824}
\and
Hidemitsu Furukawa$^c$\thanks{Tel:+81-11-706-4815} \\
$^a$Institute of Technology and Science,
The University of Tokushima, Tokushima 770-8079, Japan \\
$^b$Graduate School of Engineering,
The University of Tokushima, Tokushima 770-8079, Japan \\
$^c$Graduate School of Science,
Hokkaido University, Sapporo 060-0810, Japan}
\date{\hspace{15cm}}
\sloppy
\maketitle

\begin{abstract}
Birefringence measurements have been carried out on the Pb-doped silica hydrogels
prepared under various magnetic fields up to 5T.
The silica gels prepared at 5T were used as a medium of crystal growth of PbBr${}_2$,
whose result implied the structural anisotropy; an aligned array of crystallites
was obtained by transmission electron microscopy.
While the samples prepared at 0, 1, and 3T have no birefringence,
we found that the samples have negative birefringence on the order
of magnitude $10^{-6}$
as if the direction of the magnetic field is the optic axis of
a uniaxal crystal.
To the authors' knowledge, the birefringent silica hydrogels were obtained
by gelation under magnetic field for the first time.
Also, scanning microscopic light scattering experiments have been performed.
The results indicate that the characteristic length distribution for birefringent
samples is narrower than that for non-birefringent ones. \\[2ex]
\end{abstract}

\newpage

\section{Introduction
\label{sec:intro}}

We have found an aligned array of crystallites in gels prepared under
a magnetic field of 5T, which was used as media of the crystal growth
of PbBr${}_2$ \cite{Kaito2006JCG275,Kaito2006JCG407}. 
The crystallites were aligned with their crystallographic axis along the direction
of the magnetic field, which was applied during the preparation of the gels.
The magnetic field did not affect considerably if it was applied during
the growth of PbBr${}_2$.
It is, thus, anticipated that the magnetic field applied during the preparation
of the Pb-doped silica hydorgels made the gel structure anisotropic.
The first purpose is the identification of the anisotropy in the gel network.

Beside optimization of the condition of the crystal growth in silica gels such
as \cite{Pandita2001,Kusumoto2005,Kumari2007MRB,Kumari2007ML},
the control of the structure of silica gels is a subject of recent studies.
The control of the structure has been achieved by the selection of starting materials,
pH control of the solvent, solvent exchange during polymerization stage,
and aging \cite{Ilre,Knoblich2001,Birch2000}.

There are a lot of potential uses of the silica gels with controlled structure.
The transport of materials in the gels depends on the struscure in the gels.
Thus, for example, the crystal growth in hydorgels can be controlled by
the controlled structure in the hydrogels.
If aerogels are made by drying the controlled silica hydrogels, they keep
the structure in solution and then can be applied to a new type of
column chromatography.
If pore size is highly controlled uniformly, the filtering of the cuumn must be
improved.
If we realize anisotropy in mechanical and/or thermomechanical properties
due to the anisotropic structure of hydrogels,
those properties can be applied as a smart meterials for the sensing devices,
actuators, MEMS, and so on.

Here, we focus on the structural anisotropy in silica hydrogels and structure.
Effect of magnetic field applied during the gelation on the network
structure of polymer gels has been elucidated.
Aligned gel networks have been formed for both chemical and physical gels.
Chemically cross-linked poly(N-isopropylacrylamid) forms aligned network
structure perpendicular to magnetic field \cite{Otsuki2006}.
Physically cross-linked agarose gels also show the perpendicular alignment \cite{Yamamoto2006}.
In case side-chain group prefers the parallel alignment, polymer chain align perpendicularly
to the magnetic field because the side-chain group is basically normal to the main chain.
One of such groups contains a conjugate $\pi$ electron.
A typical example is a benzene ring, in which ring current is induced.
On the other hand, in case a group has magnetic moment, the part
including such a group preferes orient in
parallel to the magnetic field \cite{Shigekura2005}.

In the case of silica gels, the interaction of polymer chain with magnetic
field is not elucidated well.
In this paper, after presenting the results of birefringence measurement and scanning
microscopic light scattering, the discussions on the mechanism of interaction with
the magnetic field are given.

\section{Sample preparation
\label{sec:sample}}

\begin{floatingfigure}{10cm}
\begin{center}
\includegraphics[width=10cm, keepaspectratio]{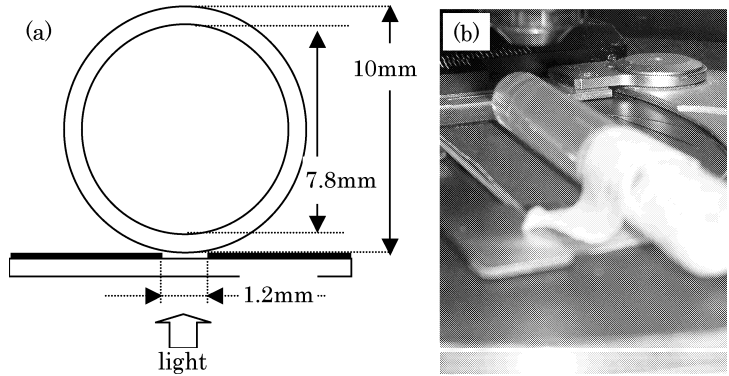}
\end{center}
\caption{\label{fig:sample}
(a) Illustration of cross-section of test tube and
(b) the sample setting at measurement of $\Delta n$.}
\end{floatingfigure}

The samples were prepared in the same way as described in \cite{Kusumoto2005},
except for the strength of the magnetic field.
Sodium metasilicate was dissolved in distilled water by starting for two hours.
After that, acetic acid and aqueous solution of Pb(NO${}_3$)${}_2$ were added and
then stirred for two hours.
This solution was settled for 7 days under various magnetic fields (0, 1, 3, and 5T)
parallel to the sample tube at a 298K.
The vessels used are cylindrical tube with inner being 7.8 mm.
The magnetic field was applied parallel to the cylindrical axis.
On the birefringence measurement a slit with the width 1.2mm to fix the geometrical
condition as illustrated in Fig.~\ref{fig:sample}; that is, we estimated the sample thickness
for this measurement to be $d$=7.71$\pm$0.09mm.

\section{Method of birefringence measurement
\label{sec:senarmont}}

\begin{floatingfigure}{7.5cm}
\begin{center}
\includegraphics[width=7.5cm, keepaspectratio]{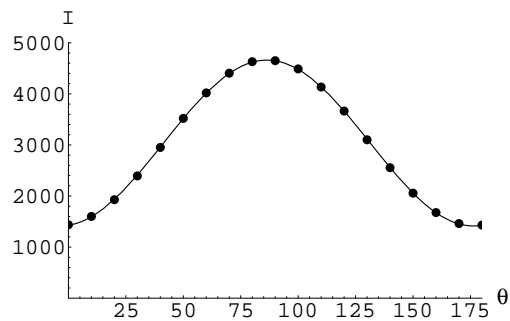}
\end{center}
\caption{\label{fig:senarmont}
A typical plot of the light intensity $I(\theta)$.}
\end{floatingfigure}

S\'{e}narmont method was employed for the birefringence measurement.
We utilized together a spectrometer to evaluate the intensity of the light transmitted
the analyzer as a function of the rotation angle of the analyzer.
Usual S\'{e}narmont method the rotation angle of the analyzer, which is a polarizer
located after quarter wave plate (S\'{e}narmont compensator), where the transmitted
light disappears is evaluated by eye.
If no sample, the transmitted light vanishes for the configuration of the crossed polarizers.
If a birefringent sample is inserted with the axis pointing the 45$^\circ$
with respect to the transmitting axis of the polarizer, the rotation angle of the analyzer
at which the transmitted light disappears changes.
Instead of detection by eye, we fit the intensity as a function of the rotation angle,
$\theta$, of the analyzer by $I=A*\cos(2\theta-\delta)+B$ to determine the retardation
$\delta=2\pi \mit\Gamma/d$ with $\mit\Gamma$ and $d$ being the optical path difference and
the thickness of the sample.
Photons were counted per 100 ms for a few minutes to obtain the intensities $I$
(averaged) and the standard deviations were also calculated.
The measure of birefringence, $\Delta n$, is obtained from $\mit\Gamma \equiv \Delta n \lambda$,
where $\lambda$ is the wavelength of light source, commonly the mercury light, whose $\lambda$
being 546nm.
Note that $\Delta n$ here is defined as $\Delta n =n_{\parallel}-n_{\perp}$ with the optic
axis being parallel ($\parallel$) to the direction of the magnetic field.
Figure~\ref{fig:senarmont} is the typical result.
In this sample the magnetic field applied was 5T and
the fitting results are $\delta=-7.22^\circ$.
Therefrom, $\Delta n$ is estimated as $-1.4032 \pm 0.15 \times 10^{-6}$.

\section{Structure characterization by dynamic light scattering
\label{sec:smils}}

Scanning microscopic light scattering (SMILS) was carried out to scan and
measure many different positions in the samples, in order to rigorously
determine a time- and space-averaged, i.e., ensemble-averaged, (auto-)
correlation function of the concentration fluctuating in the sample.
All the samples were filtered through 0.1 $\mu$m filters to avoid interference
from dust particles.
An semiconductor laser with 532 nm wavelength were used in different solutions
as the incident beam.
Measurements were made at four different angles, which were 40, 60, 90,
and 125$^\circ$, and the typical measurement time was 90 s.
The samples were maintained at a constant temperature of 30$^\circ$C throughout
the experiments.
Scanning measurement was performed at 31 points for each sample to determine
the dynamic component of the ensemble-averaged dynamic structure factor
$\Delta g_{\mbox{\scriptsize en}}^{(1)}(\tau)$ \cite{Furukawa2003}.
The determined correlation function was transformed to the distribution function
of relaxation time, $P(\tau_{\mbox{\scriptsize R}})$, numerically transformed
to relaxation-time distribution composed of one or two logarithmic Gaussian
distribution (LGD) \cite{Huang2007}.
In the present study, we simply discuss the bahavior of a main relaxation mode
of the silica hydrogels, which can be assigned to the corrective diffusion
of polymer network.

\section{Results and Discussions
\label{sec:results}}
\subsection{Birefringence
\label{sec:birfringence}}

\begin{figure}[ht]
\begin{center}
\includegraphics[width=7cm, keepaspectratio]{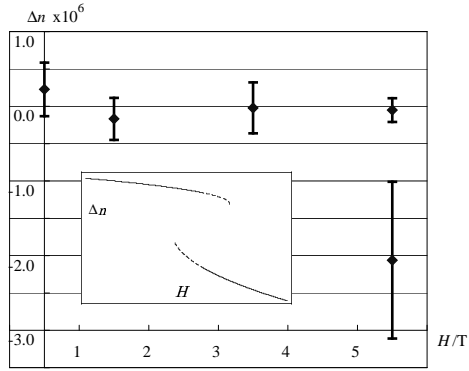}
\end{center}
\caption{\label{fig:birefringence}
The birefringence $\Delta n$ V.S. the magnetic field $H$.
The bistability is apparent at $H$=5T.
The inset is a schematic drawing for $\Delta n$-$H$ relation.}
\end{figure}

Figure~\ref{fig:birefringence}
is the results of the birefringence measurement.
In the rage of the magnetic field the samples has been classified in two classes
at $H$=5T;
one has exhibited no birefringence while a negative $\Delta n$ with the order
of the magnitude of $10^{-6}$.
It is anticipated that there exists a coexistence region where two stable states,
one of which is most stable and the other a metastable (dotted curves in the
inset of Fig.~\ref{fig:birefringence}), appear.

\subsection{Scanning microscopic light scattering
\label{sec:size}}

\begin{figure}[ht]
\begin{center}
\includegraphics[width=8cm, keepaspectratio]{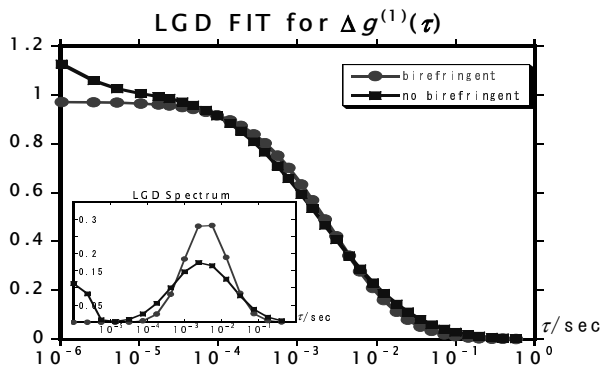}
\end{center}
\caption{\label{fig:smils}
A typical result of SMILS.
The inset is $\Delta g^{(1)}(\tau)$ with the fit by a mode distributon of the sum
of two logarythmic Gausian distribution.
The distribution of the relaxation time is obtained therefrom.
The scattering angle for these measurements is 90$^\circ$. }
\end{figure}

Typical results of SMILS are shown in Fig.~\ref{fig:smils}.
The light was input through the sample perpendicular to the sample tube.
That is, the structure in the cross sectional area perpendicular to the axes along
which the magnetic field was applied was investigated.
We saw a tendency that the distribution of $\tau$ for birefringent samples is narrower
that for non-birefringent ones.
We conjecture that the distribution of the pore size measured by SMILS directly reflects.
The optimization of the condition of measurement was difficult; the optimal combination
of the laser power and the parameter of photo multiplier was different for the scattering
angles depending of the samples.
We wish to postpone that precise determination of the pore size.

\subsection{Discussions
\label{sec:discussion}}
Here, we discuss the origin of the negative birefringence of the Pb-doped silica
gels prepared under a magnetic field.
Aforementioned we have interested in the structure of scale around several tens
or a few hundreds nm.
Pores of this scale has been reported, thought the samples have been
aerogels \cite{Wang1998}.
Closed loops must exist in the hydrogels, too.
Let us remain the existence of the dopants of Pb$^+$ and that the skeletons
made of silicon
and oxygen atoms is of full of dangling bonds.
We speculate the ring current along close loops through the mechanism similar
to that of electric conduction of the conjugated polymers \cite{Heeger1988}
and the force tending to direct those rings normal to the magnetic field.
One can readily understand the negative birefringence from this structural
anisotropy.
Other possibilities cannot be, of course, ruled out.
The wave functions or the molecular clouds, which govern the formation of
the basic structure of the gel network, in the magnetic field have been
anisotropic at the chemical reactions.

Also, narrowing of the pore size distribution can be understood consistently.
Apart from the mechanism of formation, we focus here on the orientational
distribution of the closed loops.
In case the loops align in the plane made of incident bean and the detector,
the distribution of the characteristic length become unimodal.
On the other hand, if the closed loops orients directly with respect to that
plane, the distribution must be broadened.

\section{Concluding remarks
\label{sec:conclusion}}
We have investigated the structural anisotropy of silica hydrogels prepared under
the agnatic field.
Some samples prepared under 5T exhibit negative birefringence with the optic axis
coinciding to the direction of the magnetic field.
The scanning microscopic light scattering measurments give results supporting
this structural anisotropy.
Such novel phenomena will be investigated in detail in the fufure.

\newpage
\section*{Figure captions}
\noindent
Figure~\ref{fig:sample}:
(a) Illustration of cross-section of test tube and
(b) the sample setting at measurement of $\Delta n$.\\

\noindent
Figure~\ref{fig:senarmont}:
A typical plot of the light intensity $I(\theta)$. \\

\noindent
Figure~\ref{fig:birefringence}:
The birefringence $\Delta n$ V.S. the magnetic field $H$.
The bistability is apparent at $H$=5T.
The inset is a schematic drawing for $\Delta n$-$H$ relation. \\

\noindent
Figure~\ref{fig:smils}:
A typical result of SMILS.
The inset is $g^{(1)}(\tau)$ with the fit by a mode distributon of the sum
of two logarythmic Gausian distribution.
The distribution of the relaxation time is obtained therefrom.
The scattering angle for this data is 90$^\circ$. \\


\vspace{7cm}



\newpage
\begin{thebibliography}{99} %
%
\bibitem{Kaito2006JCG275} T.~Kaito, S-i.~Yanagiya, A.~Mori, M.~Kurumada,
C.~Kaito, and T.~Inoue, J.~Cryst. Growth \textbf{289} (2006) 275-277.
%
\bibitem{Kaito2006JCG407} T.~Kaito, S-i.~Yanagiya, A.~Mori, M.~Kurumada,
C.~Kaito, and T.~Inoue, J.~Cryst.Growth \textbf{289} (2006) 407-410.

%
\bibitem{Pandita2001} S.~Pandita, V.~Hangloo, K.~K.~Bamzai, P.~N.~Kotru,
and N.~Shani, Int. J.~Inorg. Mater. \textbf{3} (2001) 675-680.
%
\bibitem{Kusumoto2005} H.~Kusumoto, T.~Kaito, S-i.~Yanagiya, A.~Mori,
and T.~Inoue, J.~Cryst. Growth \textbf{277} (2005) 536-540.
%
\bibitem{Kumari2007MRB} P.~N.~S.~Kumari, S.~Kalainathan, and N.~A.~N.~Raj,
Mater. Res. Bull. \textbf{42} (2007) 2099-2106.
%
\bibitem{Kumari2007ML} P.~N.~S.~Kumari, S.~Kalainathan, and N.~A.~N.~Raj,
Mater. Lett. \textbf{61} (2007) 4423-4425.
%
\bibitem{Ilre} R.~K.~Ilre, \textit{The Chemistry of Silica}
(Wiley-Iterscience, 1979).
%
%
%
%
%
%
\bibitem{Knoblich2001} B.~Knoblich and Th.~Gerber, J. Non-Cryst. Solid
\textbf{296} (2001) 81.
%
\bibitem{Birch2000} D.~J.~S.~Birch and C.~D.~Geddes, Phys. Rev. B.
\textbf{62} (200) 2977.
%
\bibitem{Otsuki2006} I.~Otsuki, H.~Abe and S.~Ozeki, Sci. Tech. Adv. Mater.
\textbf{7} (2006) 327-331.
%
\bibitem{Yamamoto2006} I.~Yamamoto, S.~Saito, T.~Makino, M.~Yamaguchi,
and T.~Takamas, uSci. Tech. Adv. Mater. \textbf{7} (2006) 322-236
%
\bibitem{Shigekura2005} Y.~Shgekura, Y.~M.~Chen, H.~Furukawa, T.~Kaneko,
D.~Kaneki, Y.~Osada, and J.~P. Gong, Adv. Mater. \textbf{17} (2005) 2695-2699.

\bibitem{Furukawa2003} H.~Furukawa, K.~Horie, R.~Nozaki, M.~Okada,
Phys. Rev. E \textbf{68}, (2003) 031406.
%
\bibitem{Huang2007} M.~Huang, H.~Furukawa, Y.~Tanaka, T.~Nakajima,
T.~Osada, and J.~P.~Gong, Macromolecules \textbf{40} (2007) 6658-6664.

\bibitem{Wang1998} J.~Wang, J.~Shen, B.~Zhou, Z.~Deng, L.~Zhao, L.~Zhu,
and Y.~Li, NanoStruct. Mater. \textbf{10} (1998) 909-916.
%
\bibitem{Heeger1988} A.~J.~Heeger, S.~Kivelson, J.~R.~Schrieffer, 
and W.~P.~Su, Rev. Mod. Phys. \textbf{60} (1988) 781-850.
\end{thebibliography}
\end{document}